\documentclass[twocolumn,showpacs,preprintnumbers,amsmath,amssymb,prl,superscriptaddress]{revtex4-1}

\usepackage{graphicx}
\usepackage{dcolumn}
\usepackage{bm}
\usepackage{dcolumn}
\usepackage{amsmath}
\usepackage{amssymb}
\usepackage{bbold}
\usepackage{longtable}
\usepackage{array}

\usepackage[active]{srcltx}

\newcommand{\fnm}[1]{\footnotemark[#1]}

\usepackage[usenames,dvipsnames,svgnames,table]{xcolor}

\begin{document}


\title{Impact of correlations and finite temperatures\\
  on  the anomalous Hall conductivity of 3$d$-transition-metals
} 
\author{Diemo     K{\"o}dderitzsch}    \email{dkopc@cup.uni-muenchen.de}
\address{%
  Department  Chemie,  Physikalische  Chemie,  Universit\"at  M\"unchen,
  Butenandstr.  5-13, 81377 M\"unchen, Germany\\} \author{Kristina
  Chadova} \affiliation{%
  Department  Chemie,  Physikalische  Chemie,  Universit\"at  M\"unchen,
  Butenandstr. 5-13, 81377  M\"unchen, Germany\\} \author{J\'an Min\'ar}
\affiliation{%
  Department  Chemie,  Physikalische  Chemie,  Universit\"at  M\"unchen,
  Butenandstr. 5-13,  81377 M\"unchen, Germany\\}  \author{Hubert Ebert}
\affiliation{%
  Department  Chemie,  Physikalische  Chemie,  Universit\"at  M\"unchen,
  Butenandstr. 5-13, 81377 M\"unchen, Germany\\}

\date{\today}

\begin{abstract}
  Employing the linear response Kubo formalism as implemented in a fully
  relativistic multiple-scattering Korringa-Kohn-Rostoker Green function
  method a systematic first-principles study based on density-functional
  theory  (DFT)  of  the   anomalous  Hall  conductivity  (AHC)  of  the
  3$d$-transition-metals Fe, Co and Ni is presented.  To account for the
  temperature  dependence of  the  AHC  an alloy-analogy  for  a set  of
  thermal lattice displacements acting as a scattering mechanism is used
  which   is   subsequently   solved   using  the   coherent   potential
  approximation.
  Further,  impurity scattering  has  been considered  to elucidate  the
  importance  of an  additional possible  contribution to  the  AHC that
  might be present in experiment.
  The  impact   of  correlations   beyond   the  local
  spin-density approximation  to the exchange-correlation  functional in
  DFT is studied  within the LSDA+$U$ approach.  It  is shown that both,
  the inclusion of correlations and thermal lattice vibrations, is needed
  to  give   a   material-specific  description  of   the  AHC  in
  transition-metals.   
\end{abstract}

\maketitle


The simple experiment  by Hall \cite{Hal81} driving a  current through a
ferromagnet  and  observing  the   anomalous  Hall  effect  (AHE)  as  a
transverse   voltage   has  fruitfully   spurred   the  development   of
experimental   and  theoretical  methods   dealing  with   transport  in
solids. It now stands as a paradigm for understanding related transverse
transport phenomena, like e.~g.  the spin Hall- (SHE), anomalous- and
Spin-Nernst effect which have received intense  interest in recent
years.  They all  share  a  common origin,  namely  they are  manifestly
spin-orbit driven \emph{relativistic} effects.

The AHE has for decades eluded theoretical understanding -- it took more
than 50  years until  Karplus and Luttinger  \cite{KL54} put  forward an
insight which initiated modern theories  of the AHE. They identified the
anomalous  velocity  as  an  interband  matrix element  of  the  current
operator which  is nowadays  the foundation of  semiclassical approaches
which give  a topological formulation of  the AHE in terms  of the Berry
phase of Bloch bands in  pure crystals \cite{JNM02,Sin08}. The latter is
used to define the so  called intrinsic contribution to the AHE. Already
early  on Smit  \cite*{Smi55,*Smi58} and  Berger  \cite{Ber70} discussed
other  extrinsic  origins  of   the  AHE,  namely  skew-  and  side-jump
scattering.  There  are contributions to  the AHE which fall  in neither
category, \cite{Sin08}  while it is now commonly established  to separate
the AHE into an intrinsic and a skew-scattering contribution and declare
the difference to the total AHE as side-jump \cite{NSO+10}.
Experiments   then  rely   on   scaling  mechanism   to  extract   these
contributions from the raw data.

Besides a wealth of model calculations (see the review \cite{NSO+10} and
references therein)  which are tailored  to identify general  trends but
miss  the   material  specific  aspect  a   number  of  first-principles
calculations  building on  a density-functional  theory  (DFT) framework
employing the  local spin density  (LSDA) or generalized  gradient (GGA)
approximations have  been undertaken  recently to compute  the anomalous
Hall      conductivity      (AHC)      in     the      transition-metals
(TM) \cite{WVYS07,WFS+11,FG11}.  Almost all of them rely on the Berry
phase formulation for pure crystals  and therefore are only able to deal
with  the  intrinsic  contribution.   Boltzmann-transport  theory  based
formulations have been used in the context of the SHE \cite{GFZM10} to
compute  the  skew-scattering  contributions  in the  dilute  limit  for
alloys.  Covering the whole  concentration range of alloys and including
all contributions to the  AHE has recently been done \cite{LKE10b,TKD12}
on the basis of a Kubo-St\v{r}eda formulation \cite{Str82,CB01a}.

The role of correlations in  the electronic structure of the 3$d$-TM has
only  very   recently  been  addressed   in  the  context  of   the  AHE
\cite{WFS+11,FG11}.  Employing  the LSDA  turns out to  give unfavorable
agreement  with experiment, and  with the  AHE being  a property  of the
Fermi-surface \cite{Hal04,WVYS07} it became clear that the LSDA does not
supply the proper bandstructure.  This is demonstrated in particular for
the case  of Ni (see  also table \ref{table:AHE-TM}) where  the LSDA/GGA
strongly  overestimates  the  magnitude   of  the  AHC.   Employing  the
LSDA/GGA+$U$ remedies this problem, by moving down $d$-bands relative to
the Fermi-energy ($E_F$), thereby making the $X_{2}$ hole pocket present
in LSDA/GGA  disappear.
%


A further important aspect of the AHE which is addressed in experimental
studies  but rarely  in  theoretical considerations  is the  temperature
dependence of the AHE. For the pure 3$d$-systems measurements of the AHE
are typically  done on commercially  available specimen or  thick layers
grown on a substrate  \cite{MAF+07,YTJX12,SOT09} and the temperature is
changed  in  order  to  vary  the resistivity.   The  latter  makes  the
discussion of the temperature dependence very delicate when trying to
disentangle different mechanisms and contributions to the AHE (inelastic
scattering,  scattering  by  phonons/magnons,  etc.). It  is  advocated,
however,  as an \emph{empirical  fact} \cite{NSO+10,SN12}  that inelastic
scattering processes suppress the skew-scattering at higher temperatures
with the intrinsic and side-jump (see however remark  above)
contributions  dominantly prevailing.   This then  again  is  used to
experimentally analyze the AHE.  Recently model calculations \cite{SN12}
studied  the  role  of  inelastic  scattering  by  phonons  employing  a
Kubo-formalism  and  introducing   a  phenomenological  scattering  rate
$\gamma$ as the imaginary part  of the self-energy.  To our knowledge no
first-principles  approach  has  been   used so far  to  deal  with  the
temperature dependence of AHE in 3$d$-TM.

In this letter  we present a generally applicable  formalism and results
of a first-principles approach for calculating the anomalous Hall
conductivity of  transition metals and  their alloys.  We show  that the
inclusion  of  both finite  temperature  \emph{and} correlation  effects
leads to a unified material  specific description of these systems.


As the  AHE is  inherently a relativistic  phenomenon we choose  to work
within  a fully  relativistic approach  employing the   Kohn-Sham-Dirac
equation as formulated in spin-polarized-DFT employing  the Hamiltonian:
\begin{equation}
  \label{eq:Dirac-Hamiltonian}
  {\cal H}_D =   -ic \vec{\mbox{\boldmath{$\alpha$}}} \cdot
  \vec {\mbox{\boldmath{$\nabla$}}}
  + mc^2\beta +
  \bar{V}_{KS}(\vec{r})
  +\beta\vec{\mbox{\boldmath{$\Sigma$}}}\cdot \vec{B}_{xc} (\vec{r}) \, ,
\end{equation}
with   $\bar  V_{KS}$   and  $\vec{B}_{xc}$   being the  spin-averaged  and
spin-dependent part of the one-particle potential, respectively, and the
relativistic  matrices  $\vec{\mbox{\boldmath{$\alpha$}}}$, $\beta$  and
$\vec{\mbox{\boldmath{$\Sigma$}}}$    having     the    usual    meaning
 \cite{Ros57,Ros61,Esc96}.
This  has the  important  advantage \cite{CB01a}  that  disorder can  be
treated  elegantly without  making recourse  to a  Pauli  approach which
poses difficulties in calculating the vertex corrections.
To determine longitudinal and  transverse components of the conductivity
tensor a  natural starting point  is the linear response  Kubo framework
which also can  be used to derive the  Berry phase related semiclassical
approach \cite{JNM02,SMJ+07,Sin08}.  The Kubo approach 
has important advantages as
compared  to  the  latter.   It  allows  straight-forwardly  to  include
disorder, therefore  not only being  able to describe pure  systems, but
also  alloys of  the full  concentration range  including  intrinsic and
extrinsic contributions to  the AHE \cite{LKE10b,TKD12}.  Further making
use of  an alloy  analogy model (see  below) finite temperatures  can be
accounted for.    It also allows to include
correlations  beyond LSDA  in  the framework  of  LSDA+$U$ or  LSDA+DMFT
\cite{MCE+05,Min11}.  For cubic and hexagonal
systems with the magnetization pointing
along the $\hat e_{z}$-direction, the AHE is given \cite{Str82,CB01a} by
the off-diagonal tensor element $\sigma_{\rm yx}=-\sigma_{\rm xy}$ of: 
\begin{eqnarray}
\label{eq:Kubo-streda}
\sigma_{\rm \mu\nu}
&
=
&
\frac{\hbar }{4\pi N\Omega}
{\rm Trace}\,\big\langle \hat{j}_{\rm \mu} (G^+-G^-)
\hat{j}_{\rm \nu}  G^-
\nonumber
\\
&&
\qquad \qquad \quad
-  \hat{j}_{\rm \mu} G^+\hat{j}_{\rm \nu}(G^+-G^-)\big\rangle_{\rm c}
\nonumber
\\
&&
+ \frac{|e|}{4\pi i N\Omega} {\rm Trace}\,
\big\langle (G^+-G^-)(\hat{r}_{\rm \mu}\hat{j}_{\rm \nu}
- \hat{r}_{\rm \nu}\hat{j}_{\rm \mu}) \big\rangle_{\rm c}
\label{eq:bru}
\; ,
\end{eqnarray}
with  the  relativistic current  operator $\hat{\vec  j} =  -  |e| c
\vec{\mbox{\boldmath{$\alpha$}}}$   and  the  electronic   retarded  and
advanced  Green  functions  $G^{\pm}$  which  in the  framework  of  the
presented KKR  approach are given in a  relativistic multiple scattering
representation \cite{EKM11}.    The   angular   brackets  denote   a
configurational  average which here  is carried  out using  the coherent
potential approximation (CPA) which allows to include vertex corrections
(vc),  which  are  of   utter  importance  for  the  \emph{quantitative}
determination of both the longitudinal and transversal conductivity in
alloys.  As  has been  argued and also  shown \cite{NSO+10,LKE10b,TKD12}
calculations  omitting  the  vc's  give  the  intrinsic  AHC.   Thereby,
subtracting the latter from the AHC obtained  from the value including
the vc's the extrinsic part can be extracted.

\begin{figure}
  \begin{center}
    \includegraphics[height=5cm,clip]{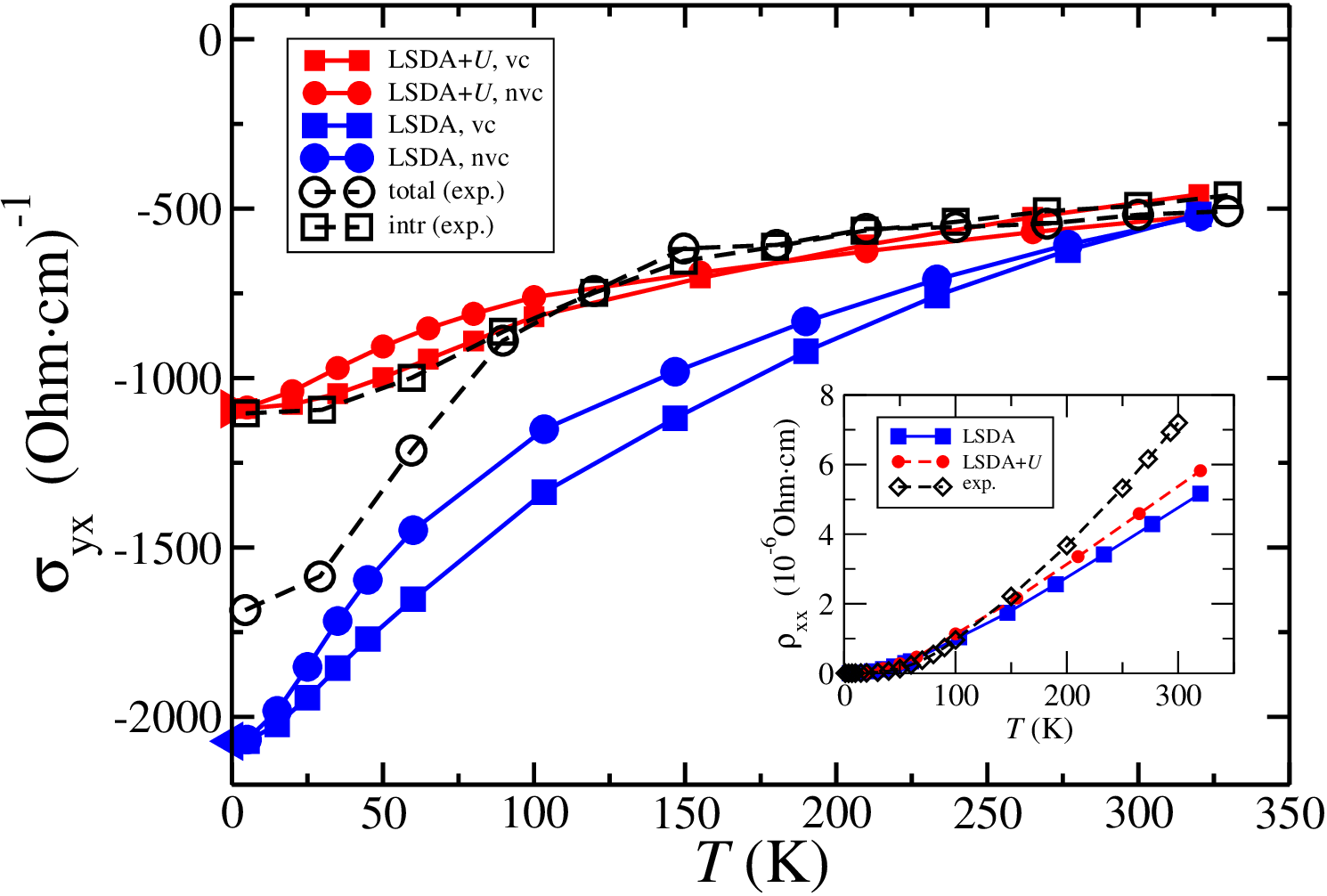}
    \caption{\label{plot:Ni-sigma-T}(Color   online)   The   temperature
      dependence  of  the  AHC  of  Ni. Shown  are  theoretical  results
      obtained  by LSDA,  LSDA+$U$  both including (vc)
      and excluding (nvc) vertex corrections 
      and experimental
      data \cite{YTJX12}. Triangle symbols denote the
      zero  temperature  values  intrinsic   AHE  values  for  LSDA  and
      LSDA+$U$,   respectively.   The   inset  shows   the  longitudinal
      $\rho_{xx}$ component of the resistivity as compared to experiment
     \cite{HAW+83}.}
  \end{center}
\end{figure}

\begin{figure}
  \begin{center}
    \includegraphics[height=5cm,clip]{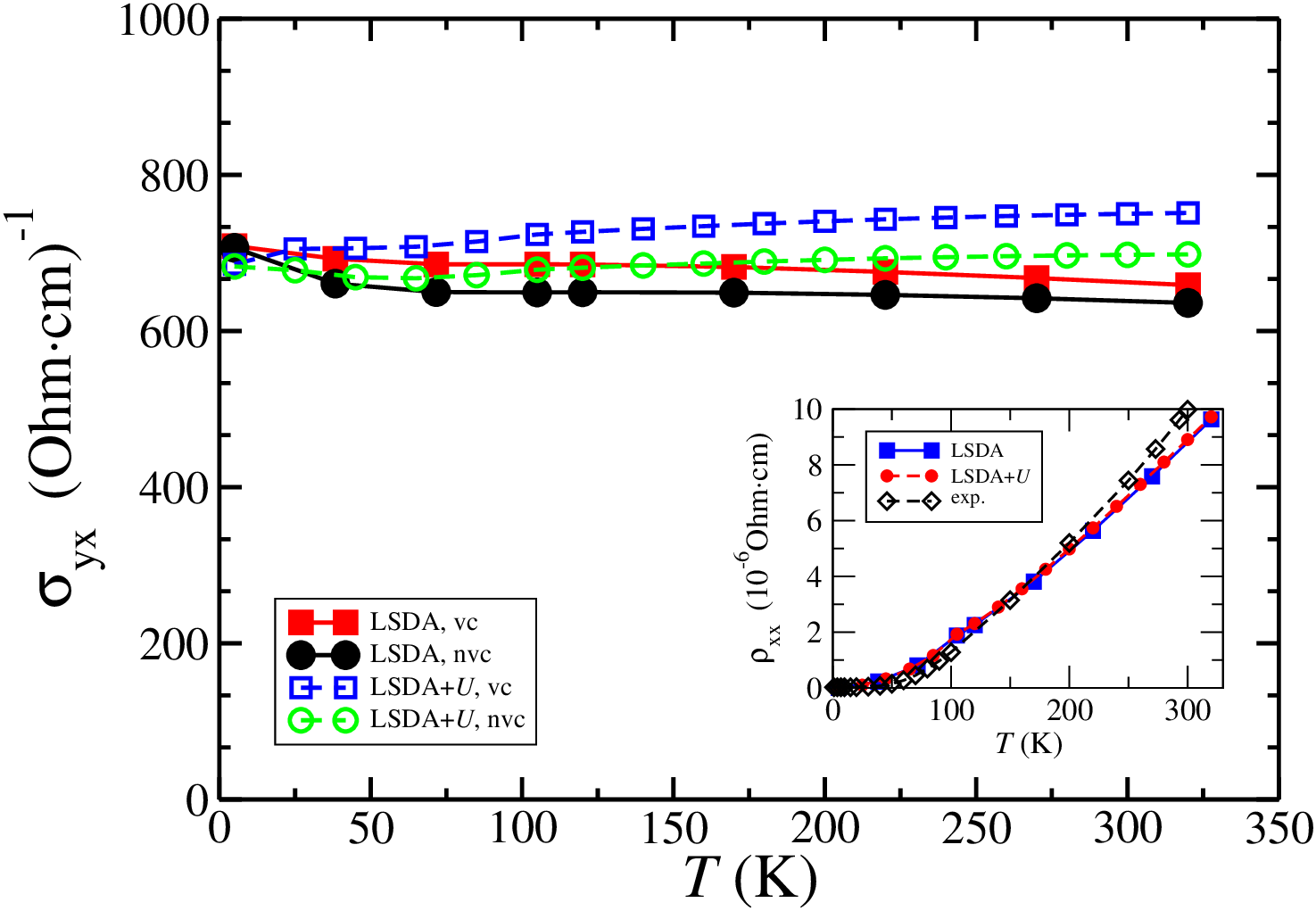}
    \caption{\label{plot:Fe-sigma}(Color    online)    The   temperature
      dependence  of  the  AHC  of Fe  as
      calculated in LSDA and LSDA+$U$. The   insets  show   the  longitudinal
      $\rho_{xx}$ component of the resistivity as compared to experiment
     \cite{HAW+83}.}
  \end{center}
\end{figure}

Several  sources  of electron  scattering  at  finite temperatures  will
determine the $T$-dependence of  the AHE.  We neglect the redistribution
of states due to finite  temperature in the electronic subsystem as well
as electron-magnon interaction  and consider as
a  dominant  effect  only   thermal  lattice  vibrations.   
 To include the latter  as a source of electron scattering
one could  generalize Eq.~(\ref{eq:Kubo-streda}) to  finite temperatures
by   including  the  electron-phonon   self-energy  $\Sigma_{\mbox{\tiny
    el-ph}}$  when  calculating the  Greens  function $G^{\pm}$.   This,
however, is computationally very expensive.  Therefore the consideration
is restricted  to elastic scattering  processes by using  a quasi-static
representation  of the  thermal displacements  of the  atoms  from their
equilibrium positions  as has  already been used successfully by the authors  in the
theory  of  Gilbert-damping \cite{EMKK11}.
Treating  each displaced  atom  as  an alloy  partner,  we introduce  an
alloy-analogy model to average over a discrete set of displacements that
is chosen to reproduce the thermal root mean square average displacement
$\sqrt{\langle u^2\rangle_T}$  for a given  temperature $T $.   This was
chosen       according      to      ${\langle       u^2\rangle_T}      =
\frac{1}{4}\frac{3h^2}{\pi^2mk\Theta_D}[\frac{\Phi(\Theta_D/T)}{\Theta_D/T}
+\frac{1}{4}]$  with  $\Phi(\Theta_D/T)$  the  Debye function,  $h$  the
Planck  constant, $k$ the  Boltzmann constant  and $\Theta_D$  the Debye
temperature \cite{GMMP83}.  Ignoring the zero temperature term $1/4$ and
assuming a  frozen potential for the  atoms, the situation  can be dealt
with in  full analogy  to the treatment  of disordered  alloys described
above.
To assess  the presented  approach the
diagonal resistivities (i.~e.  $\rho_{xx}$) were calculated for the 3$d$-TM
making use  of the Kubo-Greenwood  expression for the symmetric part  of the
conductivity tensor \cite{But86,BBVW91,BEWV94}. 
The results are shown as   insets   in
Figs.~\ref{plot:Ni-sigma-T}  and \ref{plot:Fe-sigma}  and  compared with
experimental data taken from the literature.  As can be seen the agreement
is rather  good.  Therefore we expect the aforementioned
framework  to  be  a   reasonable  approximation  to  properly  describe
electron-phonon scattering and the 
temperature dependence of the AHE.



\begin{table*}
  \centering
  \caption{\label{table:AHE-TM}The intrinsic AHC $\sigma_{\rm yx}$ in 
    $(\Omega \, {\rm cm})^{-1}$ 
    of the ferromagnetic transition metals Fe, Co and Ni from
    \emph{first-principles} theoretical (present work compared to other)
    as well as experimental (Exp.)
    studies. The magnetization has been assumed to be oriented along the
    [001] direction.
  }
  \begin{longtable}{l>{\raggedleft}p{3cm}>{\raggedleft}p{2cm}>{\raggedleft}p{2cm}>{\raggedleft}p{3cm}}
    \hline \hline
                           & bcc-Fe                & hcp-Co             & fcc-Co            & fcc-Ni                    \tabularnewline
    \hline
    LSDA, present work     & 685\hspace{2.5ex}     & 325\hspace{2.5ex}  & 213\hspace{2.5ex} & -2062\hspace{2.5ex}       \tabularnewline
    LSDA+$U$, present work & 703\hspace{2.5ex}     & 390\hspace{2.5ex}  & 379\hspace{2.5ex} & -1092\hspace{2.5ex}       \tabularnewline
    \hline
    LSDA/GGA               & 753\fnm{1},767\fnm{2} & 477\fnm{1}         & 249\fnm{3}        & -2203\fnm{1},-2200\fnm{4} \tabularnewline
                           & 650\fnm{5}            & 481\fnm{3}         & 360\fnm{5}        & -2410\fnm{5}              \tabularnewline
    LSDA/GGA+$U$           &                       &                    &                   & -960\fnm{4},-900\fnm{2}   \tabularnewline
    \hline
    Exp.                   & 1032\fnm{6}           & 813                &                   & -646(@ RT)\fnm{7}         \tabularnewline
                           &                       &                    &                   & -1100(5K)\fnm{8}          \tabularnewline
    \hline \hline
  \end{longtable}
  \footnotetext[1]{Ref.~\onlinecite{WVYS07}.}
  \footnotetext[2]{Ref.~\onlinecite{WFS+11}.}
  \footnotetext[3]{Ref.~\onlinecite{RMS09}.}
  \footnotetext[4]{Ref.~\onlinecite{FG11}.}
  \footnotetext[5]{Ref.~\onlinecite{TKD12}.}
  \footnotetext[6]{Ref.~\onlinecite{Dhe67}.}
  \footnotetext[7]{Ref.~\onlinecite{Lav61}.}
  \footnotetext[8]{Ref.~\onlinecite{YTJX12}.}
\end{table*}

%
\begin{figure}
  \begin{center}
    \includegraphics[height=3.5cm,clip]{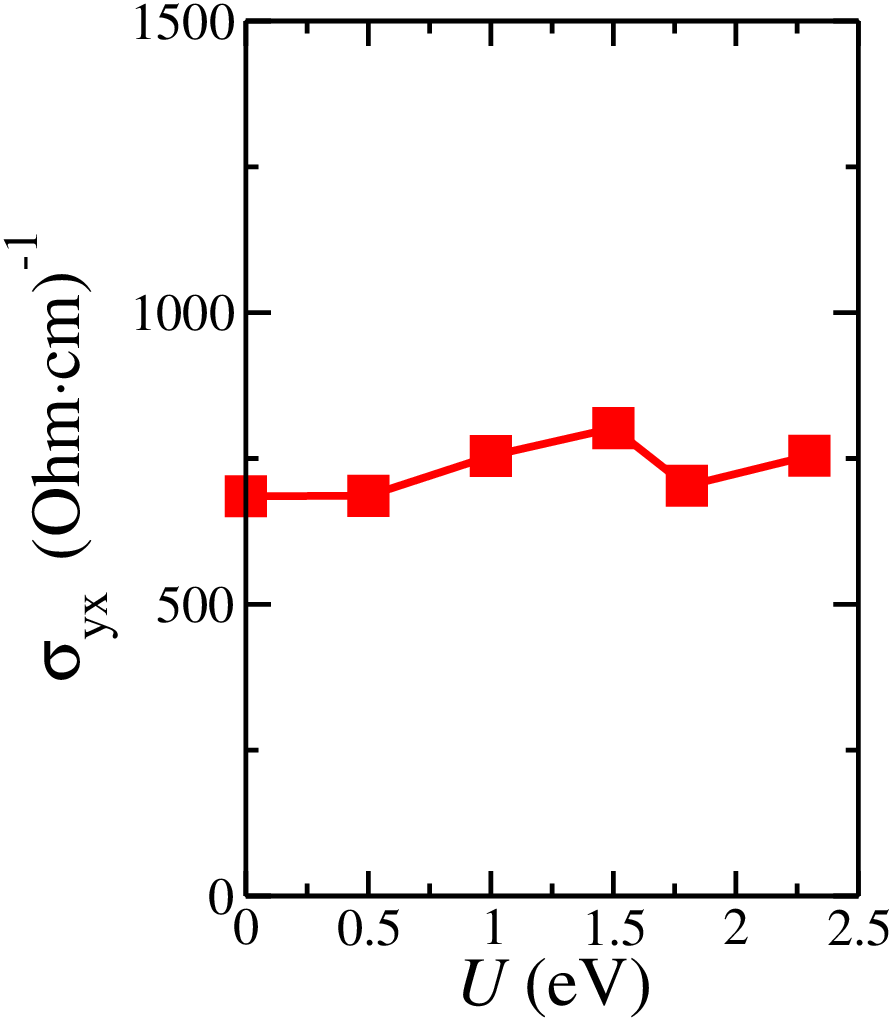}
    \includegraphics[height=3.5cm,clip]{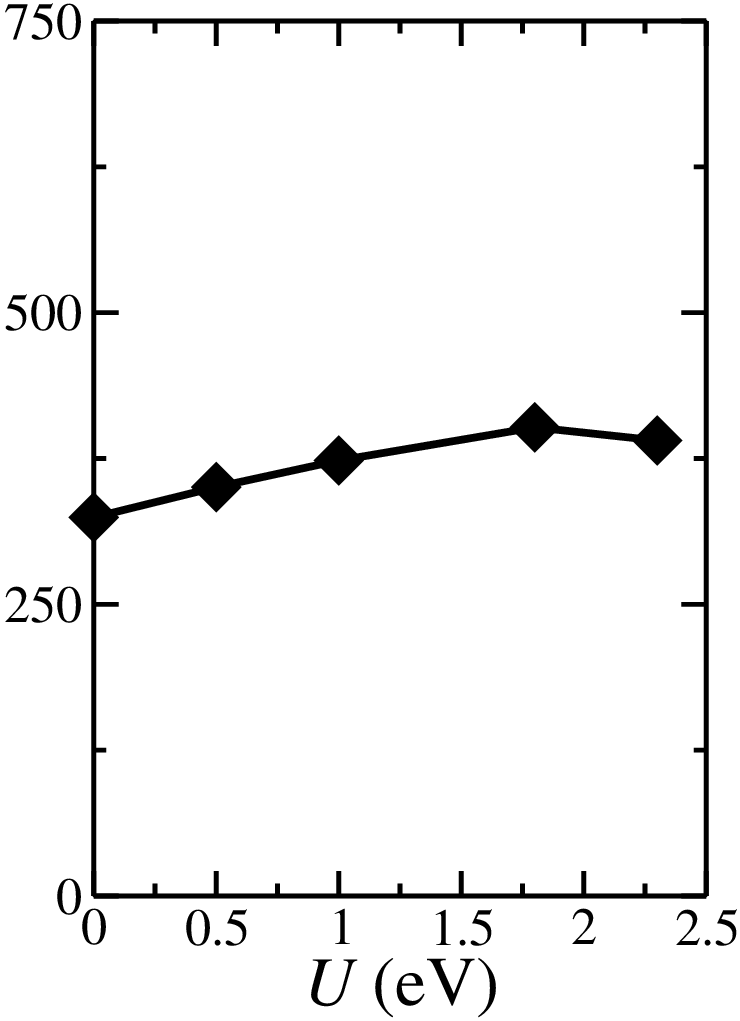}
    \includegraphics[height=3.5cm,clip]{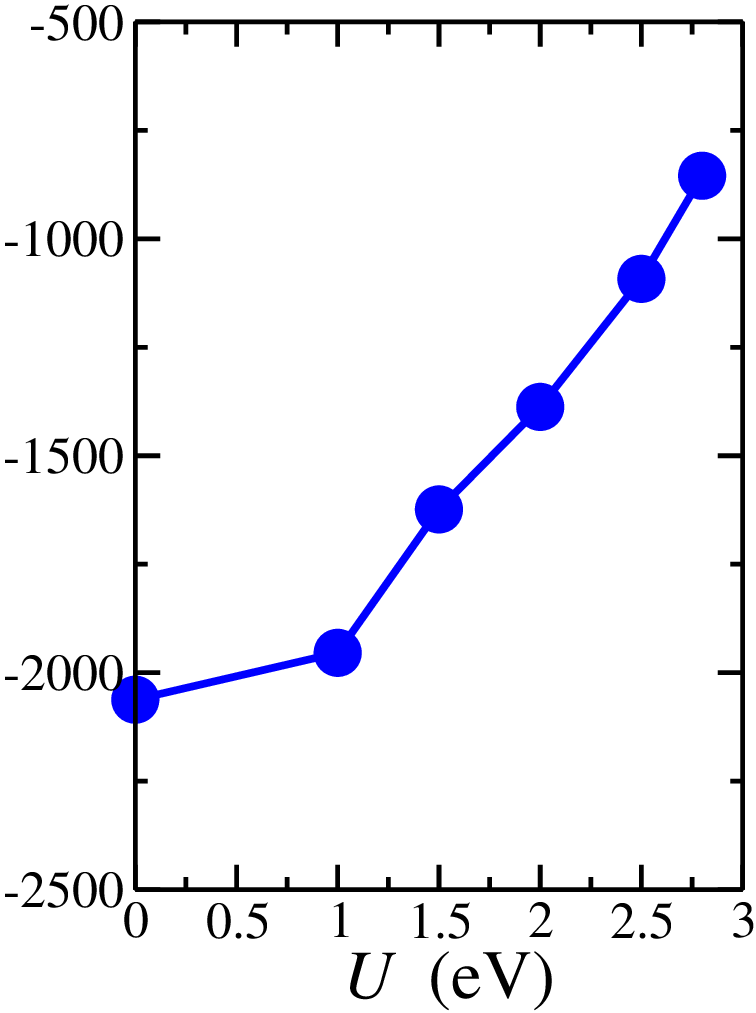}
    \caption{\label{plot:TM-sigma-U}(Color  online) The dependence  of the
      AHC ($T$=0 K) bcc-Fe, hcp-Co and fcc-Ni
      as a function of the $U$-value in the LSDA+$U$
      calculation.} 
  \end{center}
\end{figure}
%

To study the impact of correlations (beyond LSDA) LSDA+$U$ \footnote{The implementation  of the  LSDA+DMFT  in the  KKR framework  is
  reviewed in  \cite{Min11}. The LSDA+$U$  is obtained by  retaining the
  static  part of the  self-energy.  Around  mean field  double counting
  corrections  are employed.   The values  for  $U$ and  $J_{eff} $  are
  commonly used in the description of 3$d$-TM.  } calculations have been
performed  keeping $J_{\mbox{\tiny eff}}=0.9$eV  fixed and  scanning the
$U$-range  up to  typical values  employed  for the  3$d$-TMs.  In  Fig.
\ref{plot:TM-sigma-U} the dependence of the intrinsic AHE at $T=0 K$ for
bcc-Fe, hcp-Co  and fcc-Ni  is shown.  Whereas for Fe  and Co  only small
variations of the  AHC are observed,  
a  pronounced $U$-dependence for Ni is  seen  with the  experimentally
extracted  intrinsic value  of -1100  S/cm recovered  at a  $U$-value of
around  2.5eV   (this  value  is   also  used  in  calculation   of  the
$T$-dependence below).  Analysis shows that  this is due to a down shift
of minority 3$d$-bands w.r.t.\!\! $E_F$  and a vanishing hole pocket at the
$X_2$ point as has already been recently discussed \cite{FG11}.
In  table  \ref{table:AHE-TM}  we  show  the calculated  values  for  Fe
($U=1.8$eV),  Co  (hcp  and  fcc,  $U=2.3$eV )  and  Ni  ($U=2.5$eV)
\footnote{Note  that  in   our  previous  calculation  \cite{LKE10b}  we
  obtained for Ni a value of  -1635 S/cm which deviates by 20\% from the
  value reported  here. This was  due to an inappropriate  small setting
  for the muffin-tin radii  $r_{MT}$ (i.~e. no touching spheres) used
  in the calculations which employ the atomic-sphere approximation (ASA,
  $r_{ASA}$) for the potential construction.  The muffin-tin zero in the
  KKR calculation is  obtained by an averaging over the area between
  $r_{MT}$  and $r_{ASA}$.  For Ni  it turns  out that  the AHC  is very
  sensitive to such an inappropriate  setting. We checked this issue for
  Fe and Co and found  no such sensitivity, i.~e.  shrinking $r_{MT}$ by
  5\% only changed the AHC values by at most 2\%.}
in
comparison to other calculations as well as experiment. 

Early measurements of the AHE in Ni  report a value of -646 S/cm at room
temperature.   \cite{Lav61}  Recent   experimental   work  \cite{YTJX12}
analyzed this in more detail claiming the AHE to consist of an intrinsic
component of about -1100 S/cm and a sizable skew-scattering contribution
at low  temperatures, which both diminish at  higher temperatures albeit
with different rates.
%
%
In Fig.~\ref{plot:Ni-sigma-T}  the calculated temperature  dependence of
the  AHE in  Ni using  LSDA  and LSDA+$U$  ($U=2.5$ eV,  $J_{\mbox{\tiny
    eff}}=0.9 $ eV) as well as experimental results \cite{YTJX12} are
shown.  As  could be  expected from the  above the LSDA  result strongly
overestimates the  magnitude over  the whole temperature  range, whereas
the  LSDA+$U$  fairly well  reproduces  the  experimental result.   This
demonstrates that  both correlations beyond LSDA as  well as temperature
induced thermal vibrations combined need  to be taken into account.  The
vertex corrections due to the  the lattice vibrations have little impact
in the  low $T$-regime  and are negligible  at higher  temperatures such
that, as  seen in experiment, the intrinsic  contribution survives.  The
deviation  from the  total  AHC in  the  low $T$-range  we attribute  to
possible  
impurities that  might be present in  the sample. In contrast  to Ni the
temperature  dependence  in   Fe  (see  Fig.~\ref{plot:Fe-sigma})  is found to be
small. 
%
%

\begin{figure}
  \begin{center}
    \includegraphics[height=5cm,clip]{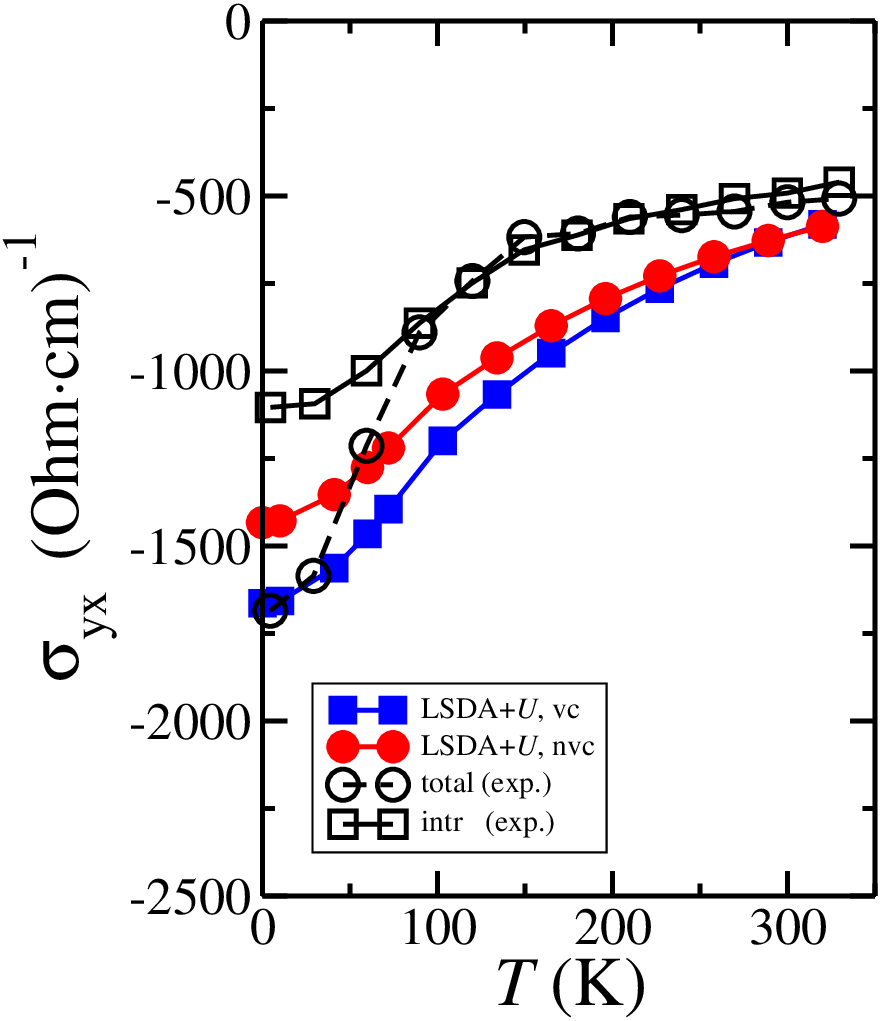}
    \includegraphics[height=5cm,clip]{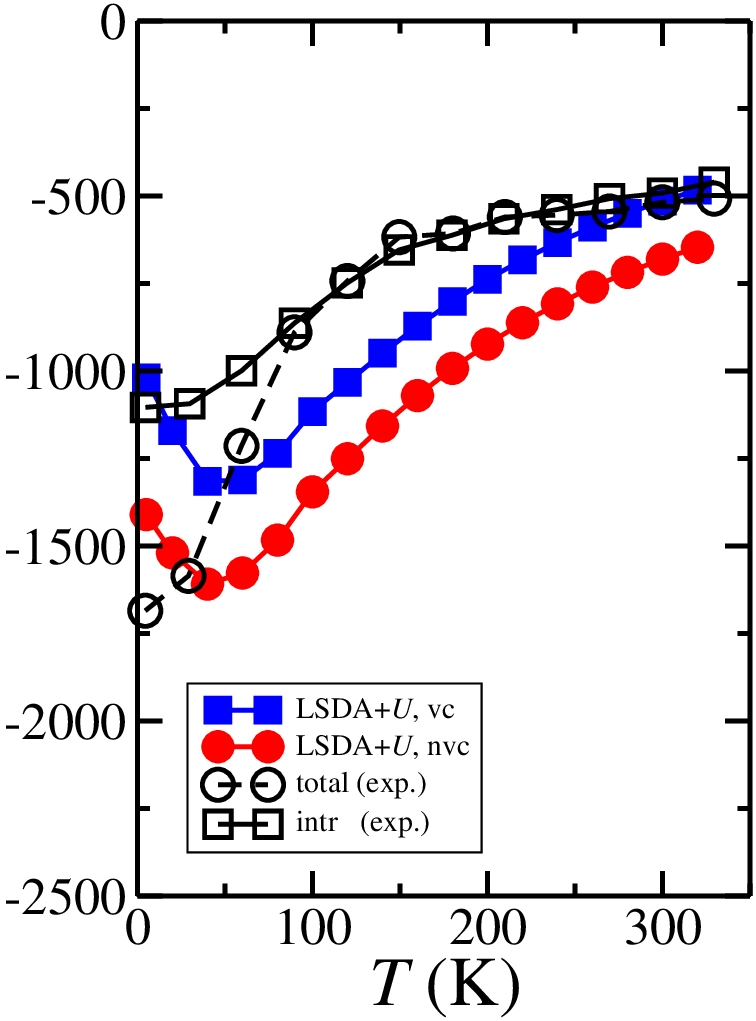}
    \caption{\label{plot:Ni-imp-sigma}(Color   online)  The  temperature
      dependence of the AHC of Ni$_{0.98}$Mg$_{0.02}$ (left) and
      Ni$_{0.98}$Fe$_{0.02}$  (right) as calculated in  LSDA+$U$   compared to
      experiment  \cite{YTJX12}. }
  \end{center}
\end{figure}

In    the     context    of    both    the    SHE     and    the    AHE,
\cite{OSN08,GFZM10,LKE10b,LGK+11}    it has
already  been shown  that  in the  dilute/super-clean  limit large  skew
scattering   contributions   can  arise   with   the   AHC  scaling   as
$\sigma_{\mbox{\tiny yx}}\propto
\sigma_{\mbox{\tiny xx}}$. To demonstrate this and put it in the context
of  the  recent  experiment  by  Ye \emph{et  al.}   \cite{YTJX12}  we
performed calculation  for Mg  impurities and Fe  impurities in  Ni.
As    can    be    seen    in
Fig~\ref{plot:Ni-imp-sigma}  the calculated  full  AHC (including  vc's)
approaches the  experimental curve for higher  temperatures.  However in
the low  temperature regime larger  deviations are visible.   Taking the
difference  between the  calculation  with vc's  to  those without  vc's
(intrinsic  values),  one  also   observes  that  the  impurity  induced
extrinsic  contribution    for Mg  shows  the  same  sign as  seen  in
experiment,  i.~e.  it  increases  the absolute  value,  whereas for  Fe
impurities  the opposite behavior  is seen.   This highlights  again the
fact  that the  skew-scattering component  in an  impurity  specific way
determines the quantitative low temperature behavior of the AHC in clean
3$d$-metals but  also that  the experimental determination  of ``clean''
systems is extremely challenging.
%
%

In  summary,  we  have  used  the  linear  response  Kubo  formalism  as
implemented  in  a  fully  relativistic  multiple-scattering  KKR  Green
function  method  to  study  systematically  from  first-principles  the
anomalous Hall  conductivity (AHC) of the  3$d$-transition-metals Fe, Co
and  Ni.   Going beyond  the  local  spin-density  approximation in  DFT
employing  the LSDA+$U$  and including  finite temperatures  by  using a
CPA-alloy analogy  for the lattice displacements  provided the necessary
means to allow for a  material specific description of the AHC. Further,
the  impact  of dilute  impurities  has  been  analyzed.  The  presented
framework is now ready to be applied to the whole concentration range of
correlated  TM-alloys.  Treating  correlations beyond  the  static limit
(LSDA+$U$) of  the LSDA+DMFT combined  with a linear  response transport
formalism is a major issue for future work.

\begin{acknowledgments}
  The authors would  like to thank the DFG  for financial support within
  the SFB 689, FOR 1346 and SPP 1538.  Discussions with Sergei Mankowsky
  are gratefully acknowledged.
\end{acknowledgments}


\begin{thebibliography}{10}

\bibitem{Hal81}
E.~H. Hall,
\newblock Phil. Mag. {\bf 12}, 157 (1881).

\bibitem{KL54}
R.~Karplus and J.~M. Luttinger,
\newblock Phys. Rev. {\bf 95}, 1154 (1954).

\bibitem{JNM02}
T.~Jungwirth, Q.~Niu, and A.~H. MacDonald,
\newblock Phys. Rev. Lett. {\bf 88}, 207208 (2002).

\bibitem{Sin08}
N.~A. Sinitsyn,
\newblock J. Phys.: Cond. Mat. {\bf 20}, 023201 (2008).

\bibitem{Smi55}
J.~Smit,
\newblock Physica {\bf 21}, 877 (1955).

\bibitem{Smi58}
J.~Smit,
\newblock Physica {\bf 24}, 39 (1958).

\bibitem{Ber70}
L.~Berger,
\newblock Phys. Rev. B {\bf 2}, 4559 (1970).

\bibitem{NSO+10}
N.~Nagaosa, J.~Sinova, S.~Onoda, A.~H. MacDonald, and N.~P. Ong,
\newblock Rev. Mod. Phys. {\bf 82}, 1539 (2010).

\bibitem{WVYS07}
X.~Wang, D.~Vanderbilt, J.~R. Yates, and I.~Souza,
\newblock Phys. Rev. B {\bf 76}, 195109 (2007).

\bibitem{WFS+11}
J.~Weischenberg, F.~Freimuth, J.~Sinova, S.~Bl\"ugel, and Y.~Mokrousov,
\newblock Phys. Rev. Lett. {\bf 107}, 106601 (2011).

\bibitem{FG11}
H.-R. Fuh and G.-Y. Guo,
\newblock Phys. Rev. B {\bf 84}, 144427 (2011).

\bibitem{GFZM10}
M.~Gradhand, D.~V. Fedorov, P.~Zahn, and I.~Mertig,
\newblock Phys. Rev. Lett. {\bf 104}, 186403 (2010).

\bibitem{LKE10b}
S.~Lowitzer, D.~K\"odderitzsch, and H.~Ebert,
\newblock Phys. Rev. Lett. {\bf 105}, 266604 (2010).

\bibitem{TKD12}
I.~Turek, J.~Kudrnovsk\'y, and V.~Drchal,
\newblock Phys. Rev. B {\bf 86}, 014405 (2012).

\bibitem{Str82}
P.~St\v{r}eda,
\newblock J. Phys. C: Solid State Phys. {\bf 15}, L717 (1982).

\bibitem{CB01a}
A.~Cr\'epieux and P.~Bruno,
\newblock Phys. Rev. B {\bf 64}, 014416 (2001).

\bibitem{Hal04}
F.~D.~M. Haldane,
\newblock Phys. Rev. Lett. {\bf 93}, 206602 (2004).

\bibitem{MAF+07}
T.~Miyasato et~al.,
\newblock Phys. Rev. Lett. {\bf 99}, 086602 (2007).

\bibitem{YTJX12}
L.~Ye, Y.~Tian, X.~Jin, and D.~Xiao,
\newblock Phys. Rev. B {\bf 85}, 220403 (2012).

\bibitem{SOT09}
Y.~Shiomi, Y.~Onose, and Y.~Tokura,
\newblock Phys. Rev. B {\bf 79}, 100404 (2009).

\bibitem{SN12}
A.~Shitade and N.~Nagaosa,
\newblock J. Phys. Soc. Japan {\bf 81}, 083704 (2012).

\bibitem{Ros57}
M.~E. Rose,
\newblock {\em Elementary Theory of Angular Momentum},
\newblock Wiley, New York, 1957.

\bibitem{Ros61}
M.~E. Rose,
\newblock {\em Relativistic Electron Theory},
\newblock Wiley, New York, 1961.

\bibitem{Esc96}
H.~Eschrig,
\newblock {\em The Fundamentals of Density Functional Theory},
\newblock B G Teubner Verlagsgesellschaft, Stuttgart, Leipzig, 1996.

\bibitem{SMJ+07}
N.~A. Sinitsyn, A.~H. MacDonald, T.~Jungwirth, V.~K. Dugaev, and J.~Sinova,
\newblock Phys. Rev. B {\bf 75}, 045315 (2007).

\bibitem{MCE+05}
J.~Min\'ar et~al.,
\newblock Nucl.\ Inst.\ Meth.\ Phys.\ Res.\ A {\bf 547}, 151 (2005).

\bibitem{Min11}
J.~Min\'{a}r,
\newblock J. Phys.: Cond. Mat. {\bf 23}, 253201 (2011).

\bibitem{EKM11}
H.~Ebert, D.~K\"odderitzsch, and J.~Min\'{a}r,
\newblock Rep. Prog. Phys. {\bf 74}, 096501 (2011).

\bibitem{EMKK11}
H.~Ebert, S.~Mankovsky, D.~K\"odderitzsch, and P.~J. Kelly,
\newblock Phys. Rev. Lett. {\bf 107}, 066603 (2011).

\bibitem{GMMP83}
E.~M. Gololobov, E.~L. Mager, Z.~V. Mezhevich, and L.~K. Pan,
\newblock phys. stat. sol. (b) {\bf 119}, K139 (1983).

\bibitem{But86}
M.~B\"uttiker,
\newblock Phys. Rev. Lett. {\bf 57}, 1761 (1986).

\bibitem{BBVW91}
J.~Banhart, R.~Bernstein, J.~Voitl\"ander, and P.~Weinberger,
\newblock Solid State Commun. {\bf 77}, 107 (1991).

\bibitem{BEWV94}
J.~Banhart, H.~Ebert, P.~Weinberger, and J.~Voitl\"ander,
\newblock Phys. Rev. B {\bf 50}, 2104 (1994).

\bibitem{RMS09}
E.~Roman, Y.~Mokrousov, and I.~Souza,
\newblock Phys. Rev. Lett. {\bf 103}, 097203 (2009).

\bibitem{Dhe67}
P.~N. Dheer,
\newblock Phys. Rev. {\bf 156}, 637 (1967).

\bibitem{Lav61}
J.~M. Lavine,
\newblock Phys. Rev. {\bf 123}, 1273 (1961).

\bibitem{Note1}
  The implementation of the LSDA+DMFT in the KKR framework is reviewed
  in \cite {Min11}. The LSDA+$U$ is obtained by retaining the static part of
  the self-energy. Around mean field double counting corrections are employed.
  The values for $U$ and $J_{eff} $ are commonly used in the description of
  3$d$-TM.

\bibitem{Note2}
Note that in our previous calculation \cite {LKE10b} we obtained for Ni a value
  of -1635 S/cm which deviates by 20\% from the value reported here. This was
  due to an inappropriate small setting for the muffin-tin radii $r_{MT}$
  (i.~e. no touching spheres) used in the calculations which employ the
  atomic-sphere approximation (ASA, $r_{ASA}$) for the potential construction.
  The muffin-tin zero in the KKR calculation is obtained by an averaging over
  the area between $r_{MT}$ and $r_{ASA}$. For Ni it turns out that the AHC is
  very sensitive to such an inappropriate setting. We checked this issue for Fe
  and Co and found no such sensitivity, i.~e. shrinking $r_{MT}$ by 5\% only
  changed the AHC values by at most 2\%.

\bibitem{HAW+83}
C.~Y. Ho et~al.,
\newblock J. Phys. Chem. Ref. Data {\bf 12}, 183 (1983).

\bibitem{OSN08}
S.~Onoda, N.~Sugimoto, and N.~Nagaosa,
\newblock Phys. Rev. B {\bf 77}, 165103 (2008).

\bibitem{LGK+11}
S.~Lowitzer et~al.,
\newblock Phys. Rev. Lett. {\bf 106}, 056601 (2011).

\end{thebibliography}

\end{document}